\documentstyle[aps,preprint]{revtex}
\begin{document}
\draft
\title{Consistent description of magnetic dipole properties\\
in transitional nuclei}
\author{Serdar Kuyucak}
\address{Department of Theoretical Physics,
Research School of Physical Sciences and Engineering,\\
Australian National University, Canberra, ACT 0200, Australia}
\author{Andrew E. Stuchbery}
\address{Department of Nuclear Physics,
Research School of Physical Sciences and Engineering,\\
Australian National University, Canberra, ACT 0200, Australia}
\date{\today}
\maketitle
\begin{abstract}
It is shown that a consistent description of magnetic dipole properties in
transitional nuclei can be obtained in the interacting boson model-2 by F-spin
breaking mechanism, which is associated with differences between the proton and
neutron deformations. In particular, the long standing anomalies observed in
the $g$-factors of the Os-Pt isotopes are resolved by a proper inclusion
of F-spin breaking.
\end{abstract}

The description of magnetic dipole ($M1$) properties in the interacting boson
model (IBM) \cite{iac87} has had a checkered career (see \cite{lip90} for a
recent review). In the original model (IBM-1 with $s$ and $d$ bosons), the
one-body $M1$ operator, being proportional to the angular momentum, results in
vanishing $M1$ transitions. Thus to explain observed $M1$ transitions one needs
at least a two-body $M1$ operator \cite{war81} whose microscopic origin is not
very clear. The discovery of the ``scissors" mode \cite{boh84} has shifted
attention to the proton-neutron version of the model (IBM-2) which provides a
more natural basis for description of $M1$ properties via the $F$-spin breaking
mechanism. $F$-spin measures the degree of symmetry between the valence protons
and neutrons and its breaking is linked to the difference between the proton
and neutron deformations \cite{lev90}. In this sense, $M1$ properties provide
complimentary information to $E2$ observables which depend on the average
deformation and are, therefore, insensitive to $F$-spin breaking.

In IBM-2, the one-body $M1$ and magnetic moment operators are given by
\begin{equation}
T(M1)=\sqrt{3/4\pi}{\bf\hat\mu}, \quad
{\bf\hat\mu} = g_\pi {\bf L}_\pi + g_\nu {\bf L}_\nu,
\end{equation}
where ${\bf L}_\rho$, $\rho=\pi,\nu$ are the angular momentum operators for
proton and neutron bosons and $g_\rho$ are the respective boson $g$-factors.
In the limit of exact $F$-spin symmetry, the $M1$ operator in (1) leads
to vanishing $M1$ transitions, and $g$-factors in a given nucleus
are constant, having the value
\begin{equation}
g(L)=(g_\pi N_\pi + g_\nu N_\nu)/N.
\end{equation}
Here $N_\pi$, $N_\nu$ denote the proton and neutron boson numbers, and
$N=N_\pi+N_\nu$. As $N_\nu$ are hole-like in the transitional
isotopes of Os and Pt,
Eq. (2) predicts an increase in $g$-factors with neutron number.
Initial IBM-2 calculations employing Eq. (2) were thought to explain
the $g$-factor variations in rare-earth nuclei reasonably well with the
bare boson $g$-factors, $g_\pi=1$, $g_\nu=0$ \cite{sam84}.
However, subsequent accurate measurements of $g$-factors at
the Australian National University and elsewhere uncovered rather large
deviations, the most conspicuous being in transitional nuclei
(see Refs.\cite{stu91,stu94} for reviews and further references).
For example, the measured $g(2^+_1)$ values for the Os isotopes increase
as predicted by Eq. (2) but the $g(2^+_2)$ values decrease, in total conflict
with it\cite{stu85}. As Eq. (2) predicts the same $g$-factor for all states,
this problem {\it cannot} be resolved by allowing arbitrary variations of
$g_\rho$ from their bare values.
In the Pt isotopes, the measured $g$-factors of the $2_1^+, 2^+_2$ and $4_1^+$
states all have similar values and remain constant with changing neutron
number.
This behaviour can be explained by Eq. (2) at the expense, however, of using
$g_\pi \simeq g_\nu \simeq 0.3$. Such large deviations of $g_\rho$ from
their bare values are not accommodated by microscopic theory,
and alternative explanations are needed.
In view of these shortcomings, attempts have been made to include
$F$-spin mixing effects via numerical diagonalization of IBM-2 Hamiltonians
\cite{stu85,zim85}. However, all of these calculations used
the parameters of Bijker {\it et al.} \cite{bij80} which were obtained by
fitting the energy levels and $E2$ transitions in Os and Pt isotopes.
Naturally, these fits are insensitive to the $F$-spin mixing needed to
describe $M1$ properties, and it is not surprising that
the above attempts did not resolve the $g$-factor discrepancies.

Recent analytic calculations using the intrinsic state formalism \cite{gin92}
and the 1/$N$ expansion method \cite{kuy89,kuy94} have provided new insight
toward the solution of this problem. The analytic expressions obtained for
$g$-factors of various bands include $F$-spin mixing effects explicitly and
have been instrumental in mapping out the parameter dependence of $M1$
properties for various $F$-spin breaking terms in the Hamiltonian. These
systematic studies have shown, in particular, that $g$-factors of ground and
excited bands respond very differently to $F$-spin breaking in the quadrupole
interaction, but show a similar behaviour toward a breaking in the one body
energies, thus suggesting that a judicious use of $F$-spin breaking could lead
to a consistent description of $M1$ properties.
The analytic formalism, though useful in pointing toward
the solution, is limited
in accuracy when applied to Os-Pt nuclei because the number of
bosons is small ($N<10$) and they have soft energy surfaces.
Thus, for accurate results, higher order terms (in $1/N$)
and band mixing contributions
must be included in the calculations. These are technically
involved and have not been performed so far.
Alternatively, one can resort to numerical diagonalization of the IBM-2
Hamiltonian with guidance from the analytic results. In  this
Letter we present the results of our numerical studies of $M1$ properties
in $^{186-192}$Os and $^{190-198}$Pt
using the computer code NPBOS \cite{npbos}.

The calculations employed the simplest IBM-2 Hamiltonian suggested by
microscopics \cite{iac87}
\begin{equation}
H=\epsilon_\pi \hat n_{d\pi} + \epsilon_\nu \hat n_{d\nu}
+ \kappa Q_\pi \cdot Q_\nu + \xi M,
\end{equation}
where $\hat n_{d\rho}$ are the $d_\rho$-boson number operators,
$M$ is the Majorana operator and $Q_\rho$ are the quadrupole operators given by
\begin{equation}
Q_\rho= [d_\rho^\dagger s_\rho + s_\rho^\dagger \tilde d_\rho]
+\chi_\rho [d_\rho^\dagger \tilde d_\rho]^{(2)}.
\end{equation}
Although other terms are often included in detailed IBM-2 studies,
Eq. (3) adequately covers the $F$-spin breaking needed to describe
$M1$ properties. In discussing $F$-spin breaking effects, it is convenient to
introduce $F$-spin scalar and vector parameters
\begin{eqnarray}
\epsilon_{\rm s}=(\epsilon_\pi+\epsilon_\nu)/2,\quad
\epsilon_{\rm v}=(\epsilon_\pi-\epsilon_\nu)/2, \nonumber\\
\chi_{\rm s}=(\chi_\pi+\chi_\nu)/2,\quad \chi_{\rm v}=(\chi_\pi-\chi_\nu)/2.
\end{eqnarray}
The $E2$ matrix elements were calculated using the the same quadrupole
operator (4) as in the Hamiltonian, with effective charges
$e_\pi=e_\nu=0.15~eb$. Bare values for the boson $g$-factors ($g_\pi=1$,
$g_\nu=0$) were employed in the $M1$ operator (1) throughout.

We first present a schematic study of $F$-spin breaking effects generated by
the two vector parameters $\epsilon_{\rm v}$ and $\chi_{\rm v}$.
Since the energies and
$E2$ transitions show little sensitivity to variations in the vector parameters
\cite{kuy89}, only the $M1$ properties are shown in Fig.~\ref{fig1}.
The effect of $F$-spin breaking on $M1$ transitions has been amply
discussed in the literature \cite{lip90} but its effect on $g$-factors has
been largely ignored until recently \cite{gin92,kuy94,wol93}.
It is clear from Fig. 1 that $g$-factors are also very sensitive to
changes in the vector parameters and, for consistency, it is
important to describe both quantities simultaneously.
Clues toward the resolution of the $g$-factor anomalies in
the Os-Pt isotopes can be
surmised from this systematic study. Specifically, the
$\epsilon_{\rm v}$ breaking
leads to a monotonic decrease in all $g$-factors as is required
in the Pt isotopes to offset the increase in theoretical values predicted by
Eq. (2). The $\chi_{\rm v}$ breaking, on the other hand, leads to a crossing of
$g$-factors of ground and $\gamma$ bands, which is precisely the behaviour
exhibited by the experimental data in $^{188-192}$Os.

The systematics of the $M1$ transitions have been included in Fig.~1
to emphasize a robust prediction of IBM-2, namely, that
the sign of $\chi_{\rm v}$
determines both the sign of $M1$ matrix element and $g(2_2^+)-g(2_1^+)$.
This prediction is consistent with the measured $g$-factors \cite{stu91}
and mixing ratio data \cite{kra80} in $^{190}$Os and $^{192}$Os
but is in conflict
with the published data in the case of $^{188}$Os.
As a by-product of the measurement of
$g$-factors in the Os isotopes \cite{stu85},
the angular correlations for the mixed $2_2^+ \to 2_1^+$ transitions were
also measured (but not published). In Fig.~\ref{fig2}, are shown the angular
correlations for the $2_2^+ \to 2_1^+$ transitions obtained in that experiment
and the resulting mixing
ratios, $\delta(E2/M1)$. The new $\delta$ values
agree well with the values in the literature, except for $^{188}$Os where the
present result ($\delta=+7.2 \pm 1.1$) has the opposite sign to that reported
previously. The angular correlation data in Fig.~2 clearly support a change
of sign in $\delta$ in $^{188}$Os, consistent with the IBM-2 prediction
for this nucleus.

In the light of the systematics discussed above, we carried out a new global
fit for $^{186-192}$Os and $^{190-198}$Pt. The parameters $\kappa$, $\xi$,
$\epsilon_{\rm s}$ and $\chi_{\rm s}$ were determined from a fit to
the energy levels and
$E2$ transitions. The first two were kept fixed in a given isotope chain while
$\epsilon_{\rm s}$ and $\chi_{\rm s}$ were slightly varied to simulate
the onset of
deformation with increasing $N_\nu$.
The vector parameters, $\epsilon_{\rm v}$ and
$\chi_{\rm v}$, were then determined from the $M1$ properties.
The parameter set
thus obtained (Table \ref{tab1}), gives a reasonable description of the
energy spectra and electromagnetic properties. The resulting $g$-factors
and $M1$ transition matrix elements are shown in Fig. \ref{fig3}.
The trends in the data are correctly reproduced and the level of agreement,
especially in the Pt isotopes, is good. We emphasize that the range of
vector parameters used in the fits (Table I) are typical of those used in
the study of $M1$ transitions in IBM-2 and that the amount of $F$-spin
admixture in the low-lying levels varies between 2-4\%, consistent with the
literature values \cite{lip90}.
Further improvement can be achieved by
fine tuning the parameters in individual nuclei and by allowing
small variations in $g_\pi$ and $g_\nu$ from their bare values.

It is of interest to note the implications of
$F$-spin mixing for the proton and neutron deformation\cite{lev90}.
To investigate
the impact of the rapid change in the vector parameters on the
proton-neutron deformations, we performed a mean field
analysis of the IBM-2 Hamiltonian (3) \cite{kuy89}. The
deformation parameters, $\beta_\pi$, $\beta_\nu$, which correspond to the mean
field amplitudes for $d_\pi$, $d_\nu$ bosons
were calculated. Since deformation in the IBM refers
to the valence nucleons only, we translate these to the more conventional
geometric model (GM) deformation ratios using the simple scaling
\begin{equation}
\left({\beta_\pi\over \beta_\nu}\right)_{GM} \approx
{N_n\over N_p}{N_\pi\over N_\nu}\left({\beta_\pi\over \beta_\nu}\right)_{IBM},
\label{gm}
\end{equation}
where $N_p$ and $N_n$ denote the number of protons and neutrons respectively.
The results of the GM deformation ratios are shown in Fig. \ref{fig4}.
As stressed in the introduction, the mean field results are not very reliable
for transitional nuclei. Further, as Eq. (\ref{gm}) is only a
qualitative relationship
between the IBM and GM, the rapid increase in the deformation ratio (Fig. 4)
is likely to be exaggerated. Nevertheless, an increasing trend is consistent
with the intuitive picture of neutrons filling the shell and hence becoming
less deformed. Recently, the proton-neutron deformation ratio was measured in
$^{165}$ Ho from pion single-charge-exchange reactions \cite{knu91}. It would
be
interesting to carry out such experiments in the Os-Pt region and
determine whether
the changes in deformation ratio are consistent with those implied by the
description of $M1$ properties in the IBM-2.

In summary, we have shown that the anomalies displayed by the $g$-factors of
the Os-Pt isotopes can be resolved in the IBM-2 by using the $F$-spin
breaking mechanism which is related to differences between the proton and
neutron deformations. A satisfactory description of the $M1$ properties has
been obtained without resorting to anomalous boson $g$-factors, effective
boson numbers, shape coexistence effects, or single particle behaviour.

This work is supported in part by the Australian Research Council.
S.K. acknowledges useful discussions with J.N. Ginocchio.

\begin{figure}
\caption{Study of $F$-spin breaking effects on $M1$ properties in $^{190}$Os.
The fixed parameters are
$\epsilon_{\rm s}=0.45$ MeV, $\chi_{\rm s}=-0.25$,
$\kappa=-0.15$ MeV, $\xi=0.17$ MeV.}
\label{fig1}
\end{figure}

\begin{figure}
\caption{Measured and fitted (solid lines) angular correlations for the
$2_2^+ \to 2_1^+$ transitions in $^{188-192}$Os.
The dashed line shows the angular correlation implied by the previously
published mixing ratio for $^{188}$Os \protect \cite{kra80}.}
\label{fig2}
\end{figure}

\begin{figure}
\caption{Experimental $g$-factors and $M1(2_2^+ \to 2_1^+)$ transition rates
compared with the present calculations. The data are from Refs.
\protect \cite{stu91,stu85,kra80,and94}.}
\label{fig3}
\end{figure}

\begin{figure}
\caption{Proton to neutron deformation ratios in Os (solid line) and Pt
(dashed line) isotopes extracted from mean field calculations.}
\label{fig4}
\end{figure}

\narrowtext
\begin{table}
\caption{Scalar and vector composition of $\epsilon$ and $\chi$
parameters in the Os-Pt isotopes.
Other parameters (in MeV):
$\kappa=-0.15$ (Os), -0.18 (Pt), and $\xi=0.17$.}
\label{tab1}
\begin{tabular}{ccccccc}
Nucleus&$N_\pi$&$N_\nu$&$\epsilon_{\rm s}$&$\epsilon_{\rm v}$
&$\chi_{\rm s}$&$\chi_{\rm v}$\\
\tableline
$^{186}$Os&3&8 & 0.32 &-0.28 & -0.32 & 0.40 \\
$^{188}$Os&3&7 & 0.35 &-0.22 & -0.28 & 0.50 \\
$^{190}$Os&3&6 & 0.40 &-0.12 & -0.25 &-0.05 \\
$^{192}$Os&3&5 & 0.40 & 0.   & -0.18 &-0.50 \\
$^{190}$Pt&2&7 & 0.45 &-0.25 &  0.18 & 0.10 \\
$^{192}$Pt&2&6 & 0.50 &-0.17 &  0.18 & 0.10 \\
$^{194}$Pt&2&5 & 0.55 &-0.07 &  0.18 & 0.05 \\
$^{196}$Pt&2&4 & 0.58 & 0.02 &  0.18 &-0.20 \\
$^{198}$Pt&2&3 & 0.58 & 0.10 &  0.18 &-0.30 \\
\end{tabular}
\end{table}

\end{document}